\documentclass[journal]{IEEEtran}

\usepackage{amsmath}

\usepackage{url}

\usepackage{amsmath}    
\usepackage{url}        
\usepackage{graphicx}   
\usepackage{float}      
\usepackage{amssymb}

\begin{document}

\title{Enhancing SQL Injection Detection and Prevention Using Generative Models}

\author{
\begin{minipage}[t]{0.256\textwidth}
    \centering
    \textbf{Naga Sai Dasari}\\
    Dept.of Computer Science,\\
    University of Reading, UK\\
    nagasai.dasari@reading.ac.uk
\end{minipage}%
\hfill
\begin{minipage}[t]{0.24\textwidth}
    \centering
    \textbf{ Atta Badii}\\
    Dept.of Computer Science,\\
    University of Reading, UK\\
    atta.badii@reading.ac.uk
\end{minipage}
\hfill
\begin{minipage}[t]{0.25\textwidth}
    \centering
    \textbf{Armin Moin}\\
    Dept.of Computer Science,\\
    University of Colorado\\
    Colorodo Springs,CO, USA\\
    amoin@uccs.edu
\end{minipage}
\hfill
\begin{minipage}[t]{0.24\textwidth}
    \centering
    \textbf{Ahmed Ashlam}\\
    Dept.of Computer Science,\\
    University of Reading, UK\\
    a.ashlam@pgr.reading.ac.uk
\end{minipage}
}
\markboth{Journal of Cybersecurity and Data Science, January 2025}%
{Dasari \MakeLowercase{\textit{et al.}}: Enhancing SQL Injection Detection and Prevention Using Generative Models}

\maketitle

\begin{abstract}
SQL Injection (SQLi) continues to pose a significant threat to the security of web applications, enabling attackers to manipulate databases and access sensitive information without authorisation. Although advancements have been made in detection techniques, traditional signature-based methods still struggle to identify sophisticated SQL injection attacks that evade predefined patterns. As SQLi attacks evolve, the need for more adaptive detection systems becomes crucial. This paper introduces an innovative approach that leverages generative models to enhance SQLi detection and prevention mechanisms. By incorporating Variational Autoencoders (VAE), Conditional Wasserstein GAN with Gradient Penalty (CWGAN-GP), and U-Net, synthetic SQL queries were generated to augment training datasets for machine learning models. The proposed method demonstrated improved accuracy in SQLi detection systems by reducing both false positives and false negatives. Extensive empirical testing further illustrated the ability of the system to adapt to evolving SQLi attack patterns, resulting in enhanced precision and robustness.
\end{abstract}


\IEEEpeerreviewmaketitle

\section{\raggedright \textbf{Introduction}}

SQL Injection (SQLi) remains one of the most critical security vulnerabilities affecting web applications today. As cyber threats evolve, attackers continuously exploit input handling weaknesses, injecting malicious SQL commands into legitimate queries. These attacks, often launched through input fields such as login forms or URL parameters, enable unauthorised access to sensitive data or, in severe cases, complete control over the database.

The Open Web Application Security Project (OWASP) continues to rank SQLi among the top security risks, reinforcing its prevalence and severity in the landscape of web vulnerabilities \cite{owasp2021}. The impact of SQLi attacks is often severe, leading to data breaches, financial loss, and reputational damage to affected organisations.

Traditional defence mechanisms, such as input validation and signature-based detection systems, have been widely employed to combat SQLi attacks. However, these methods often fall short when confronting the evolving techniques used by attackers. Signature-based systems, in particular, struggle with false positives and false negatives, especially when attackers use obfuscation or innovative variations of SQLi that deviate from known patterns \cite{anley2009, sadeghian2013, halfond2006}.

Several major challenges have made SQLi detection difficult: the lack of data diversity, static detection systems, and the high error rates of existing solutions. These limitations prevent traditional methods from adapting to the broad range of SQL Injection Attack (SQLIA) types, particularly the more complex ones.

To address these limitations, this research introduces a dynamic approach that combines synthetic data generation with advanced deep learning models. Specifically, Variational Autoencoders (VAE), U-Net, and Conditional Wasserstein GAN with Gradient Penalty (CWGAN-GP) are leveraged to generate diverse synthetic SQL data. This enriched dataset helps improve generalisation and provides better identification of both traditional and modern SQLi attacks \cite{joshi2014sql, uwagbole2017applied, farooq2021ensemble}.

The aim of this research is to enhance SQLIA detection by integrating synthetic data generation with advanced machine learning models to improve accuracy, adaptability, and overall system robustness. By preprocessing and embedding SQL queries and using synthetic data to diversify the training set, the study focuses on optimising key performance metrics such as accuracy, precision, recall, and F1-score. The remainder of this paper is organised as follows. Section II covers the literature review; Section III presents the implementation to develop the SQLi detection solution. In Section IV, results and analysis are discussed, followed by the conclusion in Section V. Finally, Section VI outlines the limitations and future scope of the work.

\section{\raggedright \textbf{Literature Review}}
\subsection{SQL Injection}

Traditional SQL Injection (SQLi) prevention methods primarily focused on fundamental coding practices such as input validation and parameterised queries, which aimed to mitigate attacks by sanitising user inputs. Although these methods were effective for basic attacks, more sophisticated techniques, such as time-based, blind, and second-order SQL injections, enabled malicious inputs to bypass traditional validation mechanisms and execute the payload at a later stage \cite{anley2009}. As SQLi threats evolved, signature-based detection systems were introduced, relying on known attack patterns to identify malicious queries in real time. However, these systems encountered significant difficulties in handling novel and obfuscated attacks that deviated from predefined patterns, resulting in high false-positive and false-negative rates \cite{sadeghian2013}. To address these challenges, rule-based systems were developed to analyse query structures more deeply. Yet, they continued to experience high false-positive rates and struggled to detect subtle attacks \cite{halfond2006}.

With the continuous advancement of SQLi techniques, behavioural detection systems were developed to identify anomalies in query behaviour. These systems aimed to detect deviations from normal query patterns but often produced high false positives, particularly in dynamic environments \cite{kiani2008}. Hybrid models that combined static code analysis with dynamic execution traces improved detection by analysing both code structure and runtime behaviour. Nonetheless, their dependency on labelled data reduced their effectiveness in real-world scenarios, where such datasets are often limited \cite{shar2013mining}. Heuristic-based systems, such as V1p3R, attempted to overcome these issues by leveraging error message feedback to adapt detection in real time, but complex and obfuscated attacks remained difficult to detect \cite{ciampa2010heuristic}.

To overcome these persistent limitations, machine learning approaches have been increasingly applied to SQLi detection. Early models, such as Naïve Bayes combined with Role-Based Access Control (RBAC), improved detection accuracy by classifying queries probabilistically. However, these models faced difficulties in handling obfuscated attacks and required manually crafted features, which limited their adaptability to novel attack patterns \cite{joshi2014sql}. Support Vector Machines (SVMs) offered further improvements by enhancing scalability and handling more complex SQLi patterns. Yet, their reliance on manual feature engineering rendered them less effective in detecting rapidly evolving attacks \cite{uwagbole2017applied}. Ensemble methods, including LightGBM and Gradient Boosting Machines (GBM), achieved high detection accuracy by combining multiple weak learners. Despite these advancements, their dependence on hand-crafted features hindered their ability to generalise to unseen queries \cite{farooq2021ensemble}.

More recently, deep learning models, such as Convolutional Neural Networks (CNNs) and Multi-Layer Perceptrons (MLPs), have pushed SQLi detection forward by automatically extracting complex patterns from SQL queries. These models reduced false positives and enhanced the detection of obfuscated attacks. However, they required large, labelled data sets and high computational resources, which limits their scalability in practical applications \cite{hasan2019detection}. SQLNN, a deep learning model that utilised TF-IDF for feature extraction, demonstrated high accuracy but faced challenges in interpretability and struggled to detect highly obfuscated queries \cite{chen2021sql}.

Despite these advancements, several challenges remain. Adapting to evolving SQLi techniques, managing limited labelled datasets, and addressing the computational costs of deep learning models continue to present significant hurdles. Traditional detection systems remain vulnerable to sophisticated attacks, while machine learning models are still dependent on manual feature engineering. These limitations have driven the exploration of adaptive solutions, such as the generation of synthetic data to augment limited datasets and improve model generalisation.

\subsection{Text Data Synthesis}

\subsubsection{Rule-based Text Synthesis}
Text data synthesis is crucial for enhancing the performance of machine learning models, offering a range of techniques to generate synthetic data. Model-based techniques, such as those explored in the work of Panagiotis et al \cite{skondras2023generating}, generate diverse data by rephrasing content while preserving its meaning, though they are computationally demanding and can introduce semantic drift. On the other hand, rule-based augmentation methods, such as synonym replacement, random insertion, and swapping \cite{awan2022augmentation}, offer computational efficiency but often fail to maintain contextual meaning, leading to distorted outputs. These limitations make rule-based methods unsuitable for complex tasks such as SQL query augmentation.

Feature-Space Augmentation, introduced in the work of Shorten et al \cite{shorten2021text}, applies transformations to latent embeddings to improve generalisation by modifying intermediate representations. Graph-Structured Augmentation preserves syntactic relationships by leveraging knowledge graphs or syntax trees, while MixUp Augmentation blends text samples and labels to expand decision boundaries and reduce overfitting. However, these methods can reduce interpretability and introduce inconsistencies, particularly in structured data such as SQL queries, where maintaining syntactic and semantic relationships is critical. Minor changes in SQL queries can disrupt the query logic, making rule-based approaches less suitable for augmenting SQL data. Consequently, model-based synthesis provides a more context-aware and accurate approach for SQL data augmentation.

\subsubsection{Model-based Text Synthesis}
Large Language Models (LLMs), as highlighted in Lovelace et al \cite{lovelace2024sample}, capture both short- and long-term dependencies, making them effective for tasks such as SQL query augmentation. However, LLMs require substantial computational resources and large data sets for training, which limits their utility in resource-constrained environments. Labs \cite{v7labs2022synthetic} discusses the use of Variational Autoencoders (VAEs) and Generative Adversarial Networks (GANs), with VAEs providing flexibility by manipulating latent space and GANs excelling in generating realistic data through adversarial training. However, GANs demand careful tuning to prevent mode collapse, which can limit their application in certain scenarios.

Transformer-based models, such as BERT, T5, and BART, utilise attention mechanisms to capture long-range dependencies and are effective for tasks such as text generation and translation. However, these models face scalability challenges when deployed at scale. Recurrent Neural Networks (RNNs), including LSTM and GRU, remain valuable for sequence modelling involving temporal dependencies but are increasingly being replaced by transformers in many scenarios. Additionally, Diffusion Models, such as Denoising Diffusion Probabilistic Models (DDPM), introduced in the work of Labs \cite{v7labs2022synthetic}, iteratively refine noisy data to improve sample quality and efficiency, though they also require significant computational resources. Seq-U-Net, introduced in the work of Stoller et al \cite{stoller2019seq}, provides a more efficient alternative for sequence modelling. By using causal convolutions, Seq-U-Net handles sequence dependencies with reduced computational costs while preserving syntactic and semantic relationships, making it particularly useful for structured tasks such as SQL query generation in resource-constrained environments.

\subsubsection{Algorithmic-based Text Synthesis}
In addition to model-based methods, algorithmic approaches such as SMOTE (Synthetic Minority Oversampling Technique) \cite{chawla2002smote} efficiently address class imbalances by generating synthetic minority class samples through interpolation. Variants such as ADASYN (Adaptive Synthetic Sampling) focus on generating samples for harder-to-learn instances, improving model performance in challenging cases \cite{kaggle2022smote}. TOMEK Links, often combined with SMOTE, refine synthetic data by removing overlapping points between majority and minority classes, enhancing classification accuracy \cite{brandt2021comparative}, \cite{zeng2016effective}. These methods ensure balanced data representation, mitigating bias and improving model generalisation.

Out of all the approaches discussed, model-based methods were selected due to their ability to maintain syntactic and semantic relationships, critical for SQL query augmentation. While SMOTE handles data imbalance, it falls short in preserving complex structures. Therefore, techniques such as  VAEs, U-Net, and GANs were chosen for their precision and reliability for structured data.

Variational Autoencoders (VAEs) have proven highly effective in reducing dimensionality and extracting features by encoding high-dimensional data into latent representations. This enables models to focus on essential features while discarding noise, making them ideal for tasks such as SQL injection detection, where computational efficiency and preserving key information are critical \cite{zhang2022deep}. Additionally, the VAE ability to handle semi-supervised learning enables efficient processing of both labelled and unlabelled data, enhancing their utility in scenarios with limited labelled datasets \cite{san2019deep}.

U-Net, initially developed for biomedical segmentation, has demonstrated its versatility in feature extraction and synthetic data generation. Its encoder-decoder architecture excels in capturing intricate features, even with minimal labelled data. This makes it particularly well-suited for generating synthetic SQL injection datasets, addressing the challenge of scarce labelled data while capturing evolving attack patterns \cite{ronneberger2015u}, \cite{zheng2020conditional}.

GANs further improve text-based data augmentation by enhancing the diversity and quality of synthetic data. Techniques such as CWGAN-GP introduce class conditioning, generating text to balance underrepresented categories, which is crucial for handling class imbalances in SQL injection datasets \cite{shorten2021text}. Additionally, models such as DP-GAN and SentiGAN promote diversity in synthetic text generation, preventing mode collapse and enhancing generalisation for text-heavy tasks \cite{xu2018diversity}, \cite{imran2022impact}.

In conclusion, model-based approaches such as VAEs, U-Net, and GANs offer advanced capabilities in generating contextually accurate and semantically rich data, making them more suitable for tasks such as SQL injection detection, despite their higher computational demands.

\section{\raggedright \textbf{Implementation}}

\begin{figure}[H]
    \centering
    \includegraphics[width=0.5\textwidth, height=0.6\textheight]{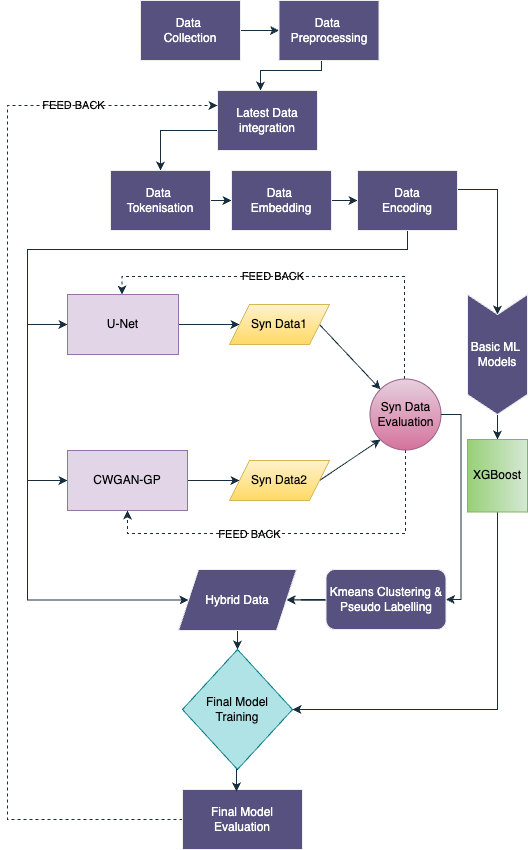}
    \caption{Pipeline Architecture}
    \label{fig:methodology}
\end{figure}

The pipeline, as illustrated in Figure \ref{fig:methodology}, details the structured stages of data processing and model implementation used in this study. The process commences with \textbf{data collection and preprocessing}, followed by \textbf{tokenisation}, \textbf{embedding}, and \textbf{encoding}. Subsequent steps include the \textbf{generation of synthetic data} using models such as U-Net, and CWGAN-GP. These synthetic datasets are then integrated with real data, resulting in a hybrid dataset, which is used for model training. The \textbf{final model evaluation} phase ensures that both real and synthetic data contribute to the detection of SQL Injection (SQLi) attacks.

\subsection{Data Collection \& Preprocessing}
The initial datasets, sourced from Kaggle (`sqli\_csv` and `Modified\_SQL\_Dataset.csv`), underwent a rigorous data cleaning process to resolve inconsistencies and remove redundant entries. These datasets were further enriched with advanced SQLi techniques such as error-based, time-based, and blind SQL injection attacks to ensure a comprehensive representation of various SQLi attack types. This integration aimed to improve the detection capabilities of the model for a broader range of SQL injection patterns, including those identified in the OWASP Top 10 A03:2021.

\subsection{Tokenisation \& Embedding}
\begin{figure}[H]
    \centering
    \includegraphics[width=\linewidth]{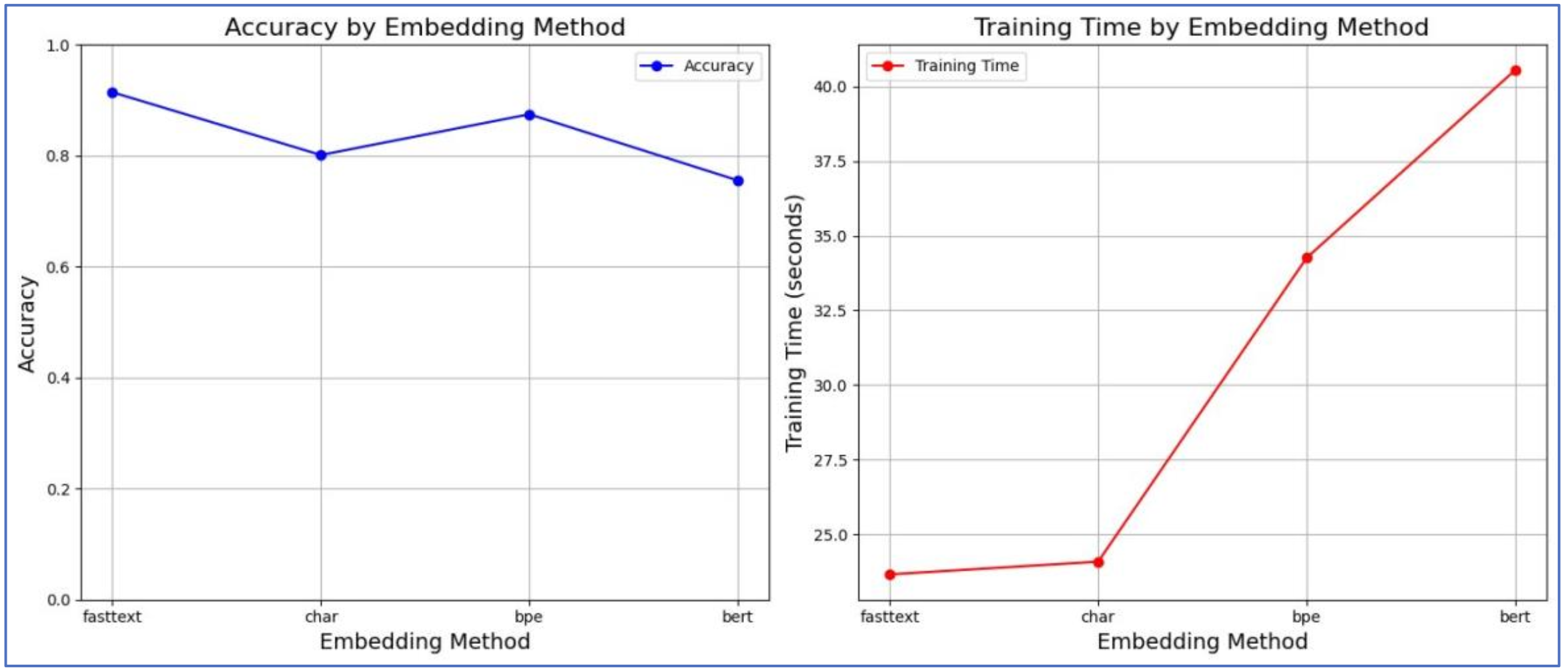}
    \caption{Accuracy and Training Time by Embedding Method}
    \label{fig:embed_accuracy}
\end{figure}

A custom tokeniser was developed to convert SQL queries into structured tokens, ensuring the capture of essential syntactic and semantic features. Various embedding methods, including \textbf{FastText}, \textbf{Character-level embeddings}, \textbf{Byte Pair Encoding (BPE)}, and \textbf{BERT}, were evaluated to determine the optimal approach. As shown in Fig. \ref{fig:embed_accuracy}, FastText emerged as the most efficient, offering a strong balance between accuracy and training time, making it the best option for transforming SQL queries into vector representations for subsequent model training.

\subsection{Feature Extraction \& Data Encoding}
\begin{figure}[H]
    \centering
    \includegraphics[width=\linewidth]{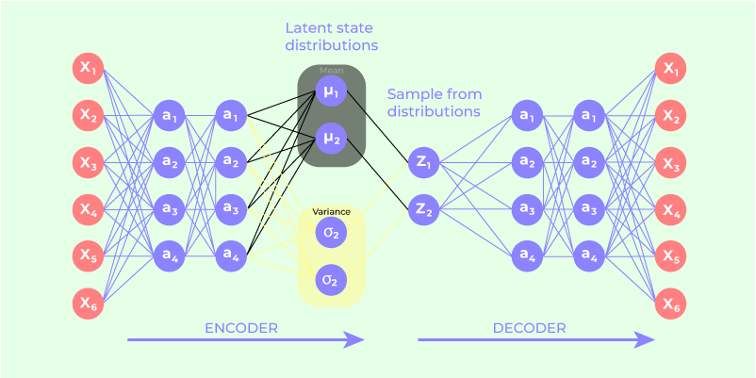}
    \caption{VAE Architecture for SQL Query Encoding}
    \label{fig:vae_architecture}
\end{figure}

The Variational Autoencoder (VAE) was employed to encode SQL queries into latent space representations, enabling effective feature extraction and dimensionality reduction. As shown in Fig. \ref{fig:vae_architecture}, the VAE consists of an encoder, which compresses the input SQL queries into latent variables characterised by mean (\(\mu\)) and variance (\(\sigma^2\)) vectors, and a decoder, which reconstructs the input data from this latent space. The initial dataset, comprising FastText embeddings with a shape of (32,336, 100, 50), was transformed into a lower-dimensional representation of shape (32,336, 448) after VAE encoding.

To enable efficient sampling from the latent space while maintaining differentiability during training, the \textbf{reparameterisation trick} was applied, as represented by the following equation:

\[
z = \mu + \epsilon \cdot \exp\left(\frac{\sigma^2}{2}\right), \quad \epsilon \sim \mathcal{N}(0,1)
\]

The VAE’s loss function combines two key components:
1. \textbf{Reconstruction loss}, which measures how accurately the decoder reconstructs the original SQL queries:
\[
\mathcal{L}_{reconstruction} = \frac{1}{N} \sum_{i=1}^{N} \|x_i - f_{\text{dec}}(z_i)\|^2
\]
2. \textbf{Kullback-Leibler (KL) divergence}, which regularises the latent space by ensuring the learned distribution is close to a unit Gaussian:
\[
\mathcal{L}_{KL} = -\frac{1}{2} \sum (1 + \log(\sigma^2) - \mu^2 - \sigma^2)
\]

The total VAE loss is expressed as:
\[
\mathcal{L}_{VAE} = \mathcal{L}_{reconstruction} + \beta \cdot \mathcal{L}_{KL}
\]
where \(\beta\) controls the trade-off between reconstruction quality and regularisation.

\begin{figure}[H]
    \centering
    \includegraphics[width=\linewidth]{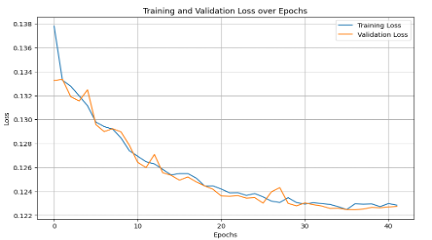}
    \caption{Training and Validation Loss during VAE Training}
    \label{fig:vae_training}
\end{figure}

Fig. \ref{fig:vae_training} illustrates the convergence of training and validation losses during VAE training, demonstrating stable learning and model generalisation.

The VAE was evaluated using several metrics, including \textbf{Mean Squared Error (MSE)}, \textbf{R² score}, and \textbf{explained variance}, ensuring that the model efficiently encoded the SQL queries while minimising reconstruction errors.

The latent representations generated by the VAE were then used as input for advanced generative models U-Net, and CWGAN-GP to generate synthetic SQL queries. Additionally, the VAE-encoded data was used to train several machine learning models, including Logistic Regression, SVM, Random Forest (RF), and XGBoost. Each model was rigorously evaluated based on accuracy, precision, recall, and F1-score to identify the best-performing baseline model. Among these, XGBoost emerged as the most effective, demonstrating superior performance in terms of classification accuracy and robustness. This baseline model was subsequently used as a reference to evaluate the quality of the synthetic data as generated by the advanced generative models.

\subsection{Synthetic Data Generation}
To enhance the diversity of the dataset, two generative models U-Net, and CWGAN-GP were utilised to generate synthetic SQL queries that closely mimic real-world SQL injection patterns. The following subsections will discuss each model in detail, outlining their architecture and adaptations for SQL query data generation.

\subsubsection{U-Net Model}

In this study, the U-Net architecture was adapted for generating synthetic SQL queries to augment the dataset used for SQL injection detection. The U-Net model was chosen due to its ability to capture both local and global dependencies, which is essential for preserving the hierarchical structure of SQL queries.

\begin{figure}[H]
    \centering
    \includegraphics[width=\linewidth]{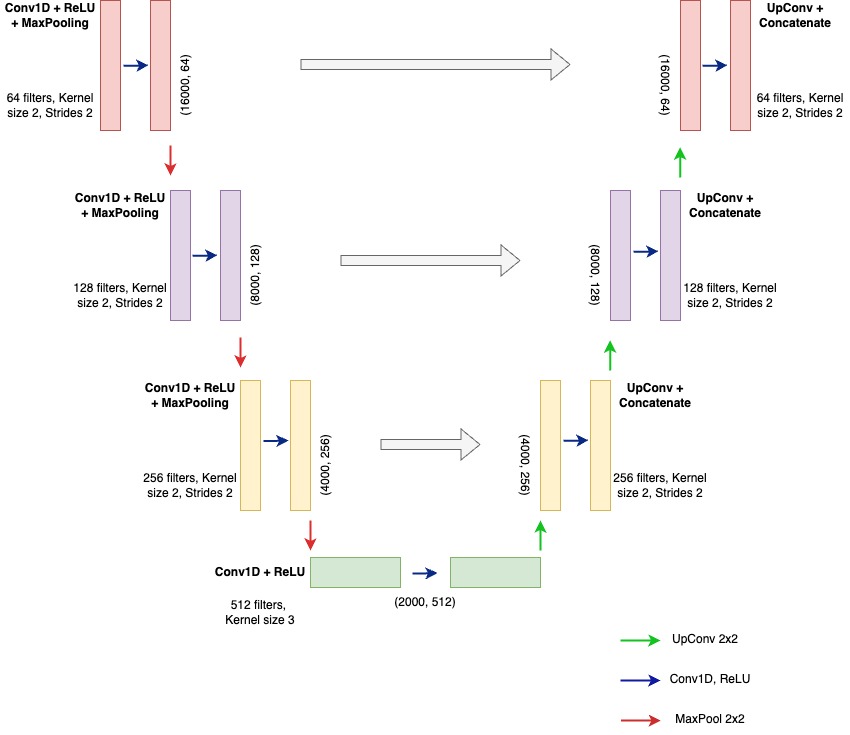}
    \caption{U-Net Model Architecture for SQL Query Generation}
    \label{fig:unet_architecture}
\end{figure}

\textbf{Model Architecture:} The U-Net model retained its core encoder-decoder architecture, but was adapted for 1D sequential data, as shown in Fig. \ref{fig:unet_architecture}. The encoder consists of convolutional layers followed by batch normalisation, ReLU activation, and max-pooling to capture abstract features and reduce dimensionality. The decoder mirrors the encoder but performs up-sampling, restoring the original sequence structure of the SQL queries while retaining critical low-level details through skip connections.

The mathematical operation for the encoder can be expressed as:
\[
f_{enc}(x) = \text{MaxPool}(\text{ReLU}(\text{BN}(\text{Conv1D}(x))))
\]
Where \( x \) is the input SQL query, \(\text{BN}\) denotes batch normalisation, and \(\text{Conv1D}\) is the 1D convolutional layer.

The decoder reconstructs the input data via:
\[
f_{dec}(x) = \text{Conv1DTranspose}(\text{Concat}(x, \text{skip}))
\]
Here, \(\text{Conv1DTranspose}\) is used for up-sampling, and \(\text{Concat}\) represents the skip connections from corresponding encoder layers.

\textbf{Hyperparameter Optimisation with Optuna:} To improve the performance of the U-Net, hyperparameter tuning was conducted using the Optuna framework. The Optuna Tree-structured Parzen Estimator (TPE) was used to explore the hyperparameter space. The optimisation aimed to minimise the Mean Squared Error (MSE) between the original and reconstructed SQL queries. The key hyperparameters optimised were:
\begin{itemize}
    \item Base Filters: Ranging from 32 to 128 filters.
    \item Learning Rate: \(1e^{-5}\) to \(1e^{-2}\).
    \item Dropout Rate: 0.1 to 0.5.
    \item Depth: 3 to 5 layers.
\end{itemize}

The best configuration included a base filter size of 704, a learning rate of \(4.61e^{-5}\), and a dropout rate of 0.03. These hyperparameters ensured a balance between model capacity and generalisation, minimising overfitting.

\textbf{Training Process:} The U-Net model was trained using the Adam optimiser, with a learning rate decay schedule that gradually reduced the learning rate. Early stopping was employed to prevent overfitting. The final loss function was defined as:
\[
L_{U-Net} = \frac{1}{N} \sum_{i=1}^{N} \left( x_i - f_{dec}(f_{enc}(x_i)) \right)^2
\]
Where \( x_i \) represents the input SQL query, and \( f_{enc} \) and \( f_{dec} \) are the encoder and decoder functions, respectively.

\begin{figure}[H]
    \centering
    \includegraphics[width=\linewidth]{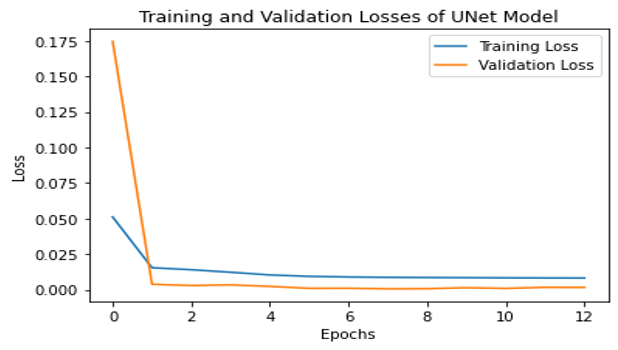}
    \caption{Training and Validation Losses of U-Net Model}
    \label{fig:unet_loss}
\end{figure}

\textbf{Evaluation Metrics:} The performance of the U-Net model was evaluated using multiple metrics, including Mean Squared Error (MSE), R² Score, Explained Variance Score (EVS), BLEU Score, Cosine Similarity, Lowenstein Distance, and Mean and Variance Differences. Additionally, Principal Component Analysis (PCA) was conducted to visually compare the real and synthetic data distributions. These metrics collectively confirm the effectiveness of U-Net in generating high-quality synthetic SQL queries.

\subsubsection{CWGAN-GP Model }

The Conditional Wasserstein Generative Adversarial Network with Gradient Penalty (CWGAN-GP) was utilised to generate synthetic SQL queries in this study. This model was chosen due to its capability to address the vanishing gradient problem and mode collapse, issues that are often encountered when training traditional GANs on complex, structured data such as SQL queries. The CWGAN-GP not only stabilises the training process but also allows the generation of SQL queries conditioned on specific labels, such as benign or malicious queries. By incorporating a gradient penalty, the model enforces Lipschitz continuity, ensuring smoother training and more realistic query generation.

\begin{figure}[H]
    \centering
    \includegraphics[width=\linewidth]{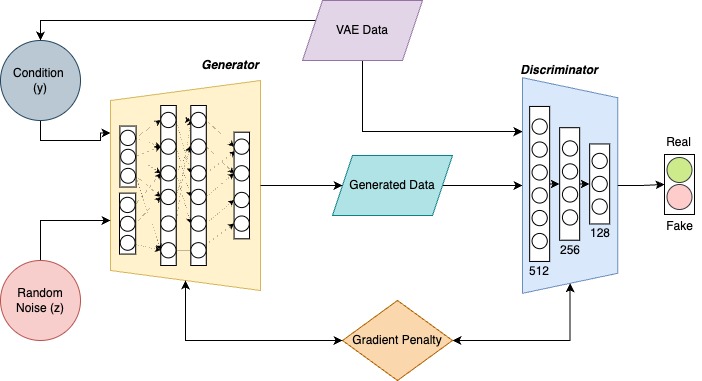}
    \caption{CWGAN-GP Architectural Design for SQL Data Generation}
    \label{fig:cwgan_architecture}
\end{figure}

\textbf{Model Architecture:} The CWGAN-GP architecture consists of two primary components, the \textit{generator} and the \textit{critic} (discriminator), as illustrated in Figure \ref{fig:cwgan_architecture}. The generator produces synthetic SQL queries by combining a random noise vector \( z \) with a one-hot encoded label \( y \), which is used to condition the output. The critic, on the other hand, evaluates both real and synthetic SQL queries, using the Wasserstein distance to distinguish between real and fake data, while a gradient penalty regularises the critic to ensure Lipschitz continuity.

\textbf{Mathematical Formulation:}

The generator \( G(z, y) \) takes a noise vector \( z \) sampled from a normal distribution and a one-hot encoded label \( y \). These inputs are concatenated and passed through several fully connected layers, which use ReLU activations to generate synthetic SQL queries. The generator can be mathematically formulated as \cite{atienza2018advanced}:

\[
G(z, y) = \text{Dense}_k(\text{ReLU}(\text{Concat}(z, y))) \to \ldots \to \text{Output Layer}
\]
where \( z \) is the latent noise vector and \( y \) is the label. The output layer produces a vector of the same dimensionality as the original SQL queries.

The critic \( D(x, y) \) receives real or generated SQL queries and their corresponding labels. The critic uses dense layers with ReLU activations to estimate the Wasserstein distance, a real-valued score that differentiates between real and fake queries. The critic is defined as Atienza \cite{atienza2018advanced}:

\[
D(x, y) = \text{Dense}_k(\text{ReLU}(\text{Concat}(x, y))) \to \ldots \to \text{Output Layer}
\]
where \( x \) is either the real or the generated SQL query.

To ensure the critic satisfies the Lipschitz constraint, a gradient penalty term is added to the loss function. The gradient penalty is computed as follows:
\[
L_{GP} = \lambda \mathbb{E} \left( ( \|\nabla_{\hat{x}} D(\hat{x}) \|_2 - 1 )^2 \right)
\]
where \( \hat{x} \) is an interpolation between real and fake data, and \( \lambda \) is a regularisation parameter controlling the contribution of the gradient penalty.

\textbf{Loss Functions:} The CWGAN-GP uses the Wasserstein loss with gradient penalty for both the generator and the critic:
\begin{itemize}
    \item \textit{Critic loss:}
    \[
    L_{Critic} = \mathbb{E}[D(x_{real})] - \mathbb{E}[D(x_{fake})] + \lambda \mathbb{E} \left( ( \|\nabla_{\hat{x}} D(\hat{x}) \|_2 - 1 )^2 \right)
    \]
    The critic maximises the difference between its evaluation of real queries \( x_{real} \) and generated queries \( x_{fake} \), while minimising the gradient penalty term to enforce stability.
    \item \textit{Generator loss:}
    \[
    L_{Generator} = - \mathbb{E}[D(x_{fake})]
    \]
    The generator minimises this loss to create synthetic queries that the critic struggles to differentiate from real queries.
\end{itemize}

\begin{figure}[H]
    \centering
    \includegraphics[width=\linewidth]{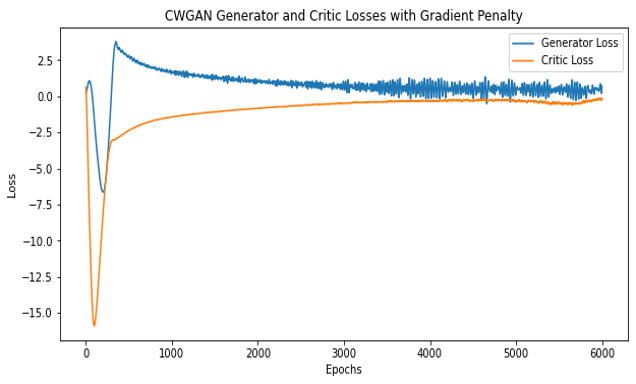}
    \caption{CWGAN-GP Generator and Critic Losses with Gradient Penalty}
    \label{fig:cwgan_loss}
\end{figure}

\textbf{Training and Optimisation:}
The training process alternates between updating the critic and the generator:
\begin{itemize}
    \item \textbf{Critic update:} The critic is updated using real and fake SQL queries, with the gradient penalty applied to enforce Lipschitz continuity. For each generator update, the critic is trained multiple times (in this case, \( n_{critic} = 2 \)) to ensure stability.
    \item \textbf{Generator update:} The generator is updated to minimise the score of the critic on the generated SQL queries.
\end{itemize}

To fine-tune the CWGAN-GP model, two approaches were employed:
\begin{itemize}
    \item \textit{Bayesian Optimisation:} Initially, Bayesian Optimisation was used to explore the hyperparameter space, resulting in minimal reconstruction loss. Hyperparameters such as the number of layers, dropout rates, and learning rate were tuned.
    \item \textit{Optuna Fine-Tuning:} After the Bayesian phase, Optuna was used to further fine-tune the model, exploring a narrower and higher-potential search space. The dynamic exploration as provided by Optuna, combined with its pruning mechanism, enabled efficient fine-tuning by halting underperforming trials early.
\end{itemize}

\textbf{Evaluation Metrics:} The performance of the CWGAN-GP model was evaluated using various metrics, including Mean Squared Error (MSE), R² score, BLEU score, Cosine similarity, and Lowenstein distance. These metrics were used to assess how closely the generated SQL queries resembled real SQL queries. Furthermore, Principal Component Analysis (PCA) was employed to visualise the overlap between real and synthetic data, confirming the CWGAN-GP model’s ability to generate realistic, high-quality SQL queries.

The results demonstrated that the CWGAN-GP model significantly improved the diversity and quality of the synthetic SQL queries, providing a robust solution for SQL injection detection systems.

\subsection{Pseudo-Labelling of Synthetic Data}
To refine the synthetic SQL data as generated by U-Net and CWGAN-GP, pseudo-labelling was employed. This method involved reducing the dimensionality of the high-dimensional data using Principal Component Analysis (PCA) and applying KMeans clustering to assign pseudo-labels.

\textbf{PCA for Dimensionality Reduction}: Principal Component Analysis (PCA) was applied to reduce the dimensions of the synthetic data to two principal components for better visual representation and easier clustering. Mathematically, the transformation can be described as:

\[
Z = XW
\]

where \( X \) represents the original high-dimensional data and \( W \) is the projection matrix consisting of the top two eigenvectors of the covariance matrix of the data. This transformation enabled a clear separation of the data into clusters, facilitating the next step of clustering and pseudo-labelling.

\textbf{KMeans Clustering for Pseudo-Labelling}: Once the data was reduced to two dimensions, KMeans clustering was performed to assign pseudo-labels. The KMeans algorithm minimised the Within-Cluster Sum of Squares (WCSS), defined as:

\[
WCSS = \sum_{i=1}^{k} \sum_{x \in C_i} \lVert x - \mu_i \rVert^2
\]

where \( k \) is the number of clusters (in this case, \( k=2 \)), \( C_i \) represents the set of points assigned to cluster \( i \), \( \mu_i \) is the centroid of cluster \( i \), and \( x \) is each data point. The KMeans algorithm assigned pseudo-labels corresponding to benign (Class 0) and malicious (Class 1) SQL queries. The labelling was based on the spread of the data: benign queries exhibited a lower spread, while malicious queries displayed a higher spread in the feature space. This distinction in the data distribution enabled the clustering algorithm to effectively separate benign from malicious queries. By leveraging these differences in data spread, the KMeans algorithm enabled the accurate classification of synthetic data into the relevant categories, facilitating its use for training machine learning models.

\subsection{Evaluation of Models on Hybrid Data}
To further enhance model performance, a hybrid dataset was created by combining real SQL data with pseudo-labelled synthetic data as generated by U-Net and CWGAN-GP models. The following steps were performed to evaluate the performance of the model:

\textbf{Hybrid Dataset Composition}: The combined dataset \( D_{\text{combined}} \) was created by mixing real data with synthetic data from U-Net and CWGAN-GP in different proportions. The combined dataset is formulated as follows:

\[
D_{\text{combined}} = D_{\text{real}} \cup (D_{\text{U-Net}} \times p_1) \cup (D_{\text{CWGAN-GP}} \times p_2)
\]

where \( D_{\text{real}} \) represents the real dataset, \( D_{\text{U-Net}} \) and \( D_{\text{CWGAN-GP}} \) are the synthetic datasets generated by U-Net and CWGAN-GP, and \( p_1 \) and \( p_2 \) are the proportions of synthetic data from each model. By adjusting \( p_1 \) and \( p_2 \), different hybrid dataset compositions were tested to optimise the training data balance between real and synthetic data.

\textbf{Cross-Validation of Dataset Combinations}: Stratified K-Fold Cross-Validation was employed to evaluate different combinations of real and synthetic data while preserving class distribution across all folds. This method ensured that the performance of the model was evaluated consistently across various splits of the data. The performance was measured using two key metrics:

\[
\text{Accuracy} = \frac{\text{True Positives} + \text{True Negatives}}{\text{Total Samples}}
\]

\[
\text{Sensitivity} = \frac{\text{True Positives}}{\text{True Positives} + \text{False Negatives}}
\]

These metrics provided insight into the ability of the model to correctly classify SQLi attacks while minimising false negatives. The cross-validation process helped identify the optimal proportion of real and synthetic data for maximising model performance, ensuring a robust balance between precision and recall.

\subsection{Final Model Evaluation}
After identifying the best dataset combination, the XGBoost classifier was trained on the combined dataset. XGBoost was selected for its high efficiency and scalability, especially in dealing with structured data such as SQL queries. The final model used logistic loss as the objective function, defined as:

\[
L_{\text{XGBoost}} = -\frac{1}{N} \sum_{i=1}^{N} \left[ y_i \log \hat{y}_i + (1 - y_i) \log (1 - \hat{y}_i) \right]
\]

where \( N \) is the total number of samples, \( y_i \) is the true label for sample \( i \), and \( \hat{y}_i \) is the predicted probability for sample \( i \). This loss function optimises the classification model by minimising the error in predicting the correct labels for both benign and malicious queries.

The trained XGBoost model was evaluated on the test set using multiple metrics, including accuracy, sensitivity, precision, recall, and F1-score. 

The results demonstrated that the combination of real and pseudo-labelled synthetic data improved the ability of the model to generalise to new, unseen SQL queries. The final XGBoost model achieved high accuracy, sensitivity, and precision, indicating its effectiveness in SQL injection detection across diverse attack types.

\section{\raggedright \textbf{Results and Analysis}}

In this section, the performance of several machine learning models trained on VAE-encoded SQL data is evaluated for detecting SQL Injection Attacks (SQLIA). The models tested include XGBoost, LightGBM, Random Forest, K-Nearest Neighbors (KNN), Neural Networks, Logistic Regression, Support Vector Classifier (SVC), and Naive Bayes. The evaluation focuses on metrics such as accuracy, precision, recall, F1-score, and sensitivity for both benign (Class 0) and malicious (Class 1) queries.

\subsection{Classification Metrics}
Several standard classification metrics were used to thoroughly assess the performance of the models. These metrics provide valuable insights into how each model predicts benign (Class 0) and malicious (Class 1) SQL queries. The following metrics were calculated for each model:

\begin{table}[H]
\centering
\caption{Classification Metrics and Formulas}
\begin{tabular}{|c|c|p{3.5 cm}|}
\hline
\textbf{Metric} & \textbf{Formula} & \textbf{Description} \\ \hline
\textbf{Accuracy} & 
$\frac{TP + TN}{TP + TN + FP + FN}$ & 
Proportion of correct predictions out of all predictions \\ \hline
\textbf{Precision} & 
$\frac{TP}{TP + FP}$ & 
Proportion of positive predictions that were correct \\ \hline
\textbf{Recall} & 
$\frac{TP}{TP + FN}$ & 
Proportion of actual positives correctly predicted (Sensitivity for Class 1) \\ \hline
\textbf{F1-Score} & 
$2 \times \frac{\text{Precision} \times \text{Recall}}{\text{Precision} + \text{Recall}}$ & 
Harmonic mean of precision and recall \\ \hline
\textbf{Sensitivity} & 
$\frac{TP_{\text{Class 1}}}{TP_{\text{Class 1}} + FN_{\text{Class 1}}}$ & 
Sensitivity is equivalent to recall for Class 1 \\ \hline
\end{tabular}
\end{table}

The performance of various machine learning algorithms was evaluated based on their ability to detect SQL Injection Attacks (SQLIA). 

\begin{figure}[H]
    \centering
    \includegraphics[width=0.5\textwidth]{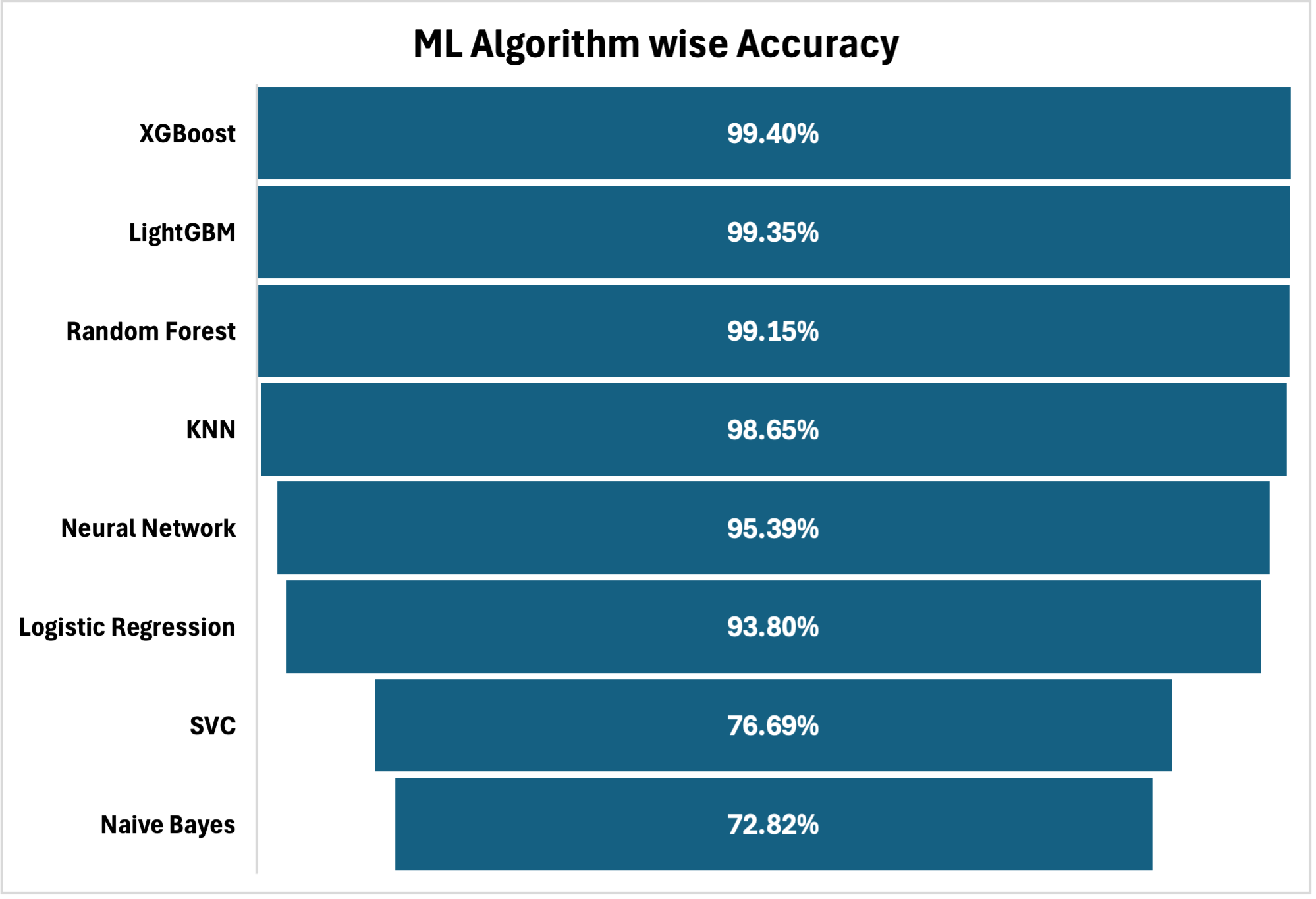}
    \caption{Comparison of Machine Learning Algorithms based on Accuracy. XGBoost outperforms others in SQLIA detection accuracy.}
    \label{fig:ML_models_accuracy}
\end{figure}

As depicted in Figure \ref{fig:ML_models_accuracy}, XGBoost achieved the highest accuracy, outperforming other models with an impressive score of 99.40\%. LightGBM followed closely with 99.35\%, while Random Forest and KNN achieved 99.15\% and 98.65\%, respectively. Neural Networks, while often strong performers in complex tasks, registered 95.39\% in this specific SQLIA detection task. Logistic Regression, Support Vector Classifier (SVC), and Naive Bayes were outperformed by the others, with Naive Bayes being the least accurate at 72.82\%.

The results underscore the suitability of XGBoost for SQLIA detection tasks, making it the preferred choice for further experiments. Its balance of speed, precision, and handling of imbalanced data makes it ideal for this application, particularly when combined with synthetic data generation techniques such as those explored in this study.

\subsection{Synthetic Data Quality Evaluation Metrics}

To assess the quality of the synthetic SQL query data as generated by U-Net and CWGAN-GP, several key metrics were utilised. These metrics are essential for ensuring the synthetic data closely aligns with real SQL queries in both structural and statistical dimensions, which is critical for SQL injection detection models.

\begin{itemize}
    \item \textbf{Mean Squared Error (MSE)}: MSE calculates the average squared difference between the real and synthetic data. A lower MSE value indicates that the generated data closely mimics the real data, reducing the risk of discrepancies in model training \cite{goldstein1999summarizing}.

    \item \textbf{R² Score}: This metric quantifies the amount of variance in the real data that is captured by the synthetic data. A high R² score ensures that the generated SQL queries retain sufficient diversity, supporting better generalisation in machine learning models.

    \item \textbf{Explained Variance Score (EVS)}: EVS measures the proportion of variance in the real dataset captured by the synthetic data. High EVS indicates that the synthetic queries adequately cover the behaviours and patterns observed in real-world SQL injection attacks \cite{goldstein1999summarizing}.

    \item \textbf{BLEU Score}: BLEU measures the token-level similarity between real and synthetic queries. A higher BLEU score reflects greater structural alignment, which is critical for maintaining the functional semantics of SQL queries \cite{goldstein1999summarizing}.

    \item \textbf{Cosine Similarity}: This metric calculates the angular similarity between vectorised representations of real and synthetic data. In SQL query generation, cosine similarity ensures that the overall semantic meaning of the queries is preserved.

    \item \textbf{Lowenstein Distance}: This computes the number of edits required to transform synthetic queries into real ones. A lower distance signifies a closer match between the real and synthetic data, which is vital for maintaining query integrity.

    \item \textbf{Mean and Variance Differences}: These metrics compare the statistical distributions of real and synthetic data by evaluating their means and variances. In SQL data generation, this ensures that the statistical properties, such as query patterns and structures, are preserved, improving the reliability of the generated data for SQL injection detection.

\end{itemize}

\textbf{Other Metrics and Justifications}

Several other metrics commonly used in text generation, such as perplexity and compression ratio, were considered but deemed unsuitable for this structured dataset:

\begin{itemize}
    \item \textbf{Perplexity}: This metric is typically used in language modelling to measure uncertainty in word predictions. However, because SQL queries are deterministic and do not involve probabilistic word choices, perplexity is not applicable in this context \cite{goldstein1999summarizing}.

    \item \textbf{Compression Ratio}: Commonly used to evaluate text summarisation, this metric measures the reduction in text length. In SQL query generation, the goal is to preserve accuracy and completeness rather than conciseness, making this metric inappropriate for the task.
\end{itemize}

By selecting the metrics most relevant to structured SQL data, this evaluation ensures that the synthetic queries generated are reliable and useful for training machine learning models to detect SQL injection attacks.

In addition, Principal Component Analysis (PCA) and K-Means Clustering were used to visually inspect the alignment between real and synthetic data distributions. These techniques provide additional insights into the structural similarities of both datasets. The combination of these metrics ensures a robust evaluation of the synthetic data's quality, supporting its use in training models for SQL Injection detection systems.

\subsection{Evaluation of Synthetic Data}
\subsubsection{U-Net Model: Results and Discussion}
The performance of the U-Net model in generating synthetic SQL queries was evaluated using the above key metrics. 

\begin{figure}[H]
    \centering
    \includegraphics[width=0.3\textwidth]{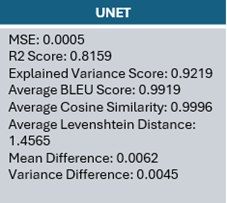}
    \caption{Performance Metrics for U-Net Model}
    \label{fig:unet_metrics}
\end{figure}

As shown in Figure \ref{fig:unet_metrics}, the U-Net model performed exceptionally well, achieving a low Mean Squared Error (MSE) of 0.0005, indicating a strong match between the real and synthetic SQL data. The model’s R² score of 0.8159 of the model confirms its effectiveness in capturing variance within the real dataset.

Other key metrics include a BLEU Score of 0.9919 and a Cosine Similarity of 0.9996, which highlight the close structural and semantic resemblance between the generated and real queries. Furthermore, the Lowenstein Distance of 1.4565 supports the high similarity between the datasets, requiring minimal changes for alignment.

\begin{figure}[H]
    \centering
    \includegraphics[width=0.5\textwidth]{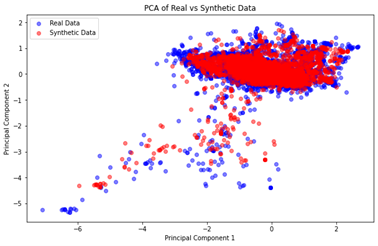}
    \caption{PCA of Real vs Synthetic Data for U-Net Model}
    \label{fig:pca_unet}
\end{figure}

The Principal Component Analysis (PCA) results in Figure \ref{fig:pca_unet} further validate the consistency of the U-Net model in generating synthetic queries that align closely with the real SQL data. Minimal distributional differences, as indicated by the Mean Difference of 0.0062 and Variance Difference of 0.0045, emphasise the strong generalisation capabilities of the model in mimicking real-world query structures.

\subsubsection{CWGAN-GP Results and Discussion}

Similar to the U-Net model, the synthetic data generated by CWGAN-GP was evaluated using key performance metrics to assess its effectiveness.

\begin{figure}[H]
    \centering
    \includegraphics[width=0.3\textwidth]{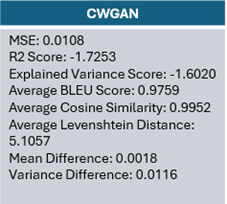}
    \caption{Performance Metrics for CWGAN-GP Model}
    \label{fig:cwgan_metrics}
\end{figure}

As illustrated in Figure \ref{fig:cwgan_metrics}, the CWGAN-GP model achieved a Mean Squared Error (MSE) of 0.0108, indicating some variance between the real and synthetic datasets. The R² score of -1.7253 and the Explained Variance Score of -1.6020 further reflect deviations from real data patterns.

Despite these deviations, the model produced a BLEU Score of 0.9759 and a Cosine Similarity of 0.9952, showing strong structural alignment between the real and synthetic SQL queries. The Lowenstein Distance of 5.1057, though higher, suggests moderate similarity in sequence structure between the generated and real data.

\begin{figure}[H]
    \centering
    \includegraphics[width=0.5\textwidth]{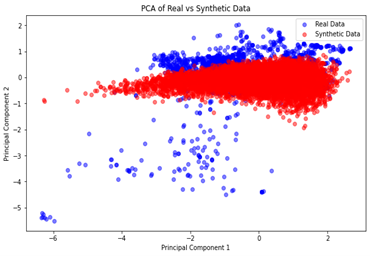}
    \caption{PCA of Real vs Synthetic Data for CWGAN-GP Model}
    \label{fig:pca_cwgan}
\end{figure}

Figure \ref{fig:pca_cwgan} shows the Principal Component Analysis (PCA), where the real and synthetic data exhibit partial overlap, with the synthetic data displaying greater dispersion. This suggests that while CWGAN-GP effectively captures overall patterns, there remains some variability.

In summary, the CWGAN-GP model demonstrates effective token-level and vector-based similarities with real data, though with some structural deviations, as evidenced by the PCA and higher Lowenstein Distance.

\subsection{Pseudo-Labelling and Clustering Results}

Following the generation of synthetic SQL data, pseudo-labelling was employed to categorise the data into distinct classes. Pseudo-labels were assigned using KMeans clustering, which organised the data based on feature similarities. The spread of the data was used to determine the labels: benign queries (Class 0) exhibited a lower spread, while malicious queries (Class 1) displayed a higher spread.

\begin{figure}[H]
    \centering
    \includegraphics[width=0.5\textwidth]{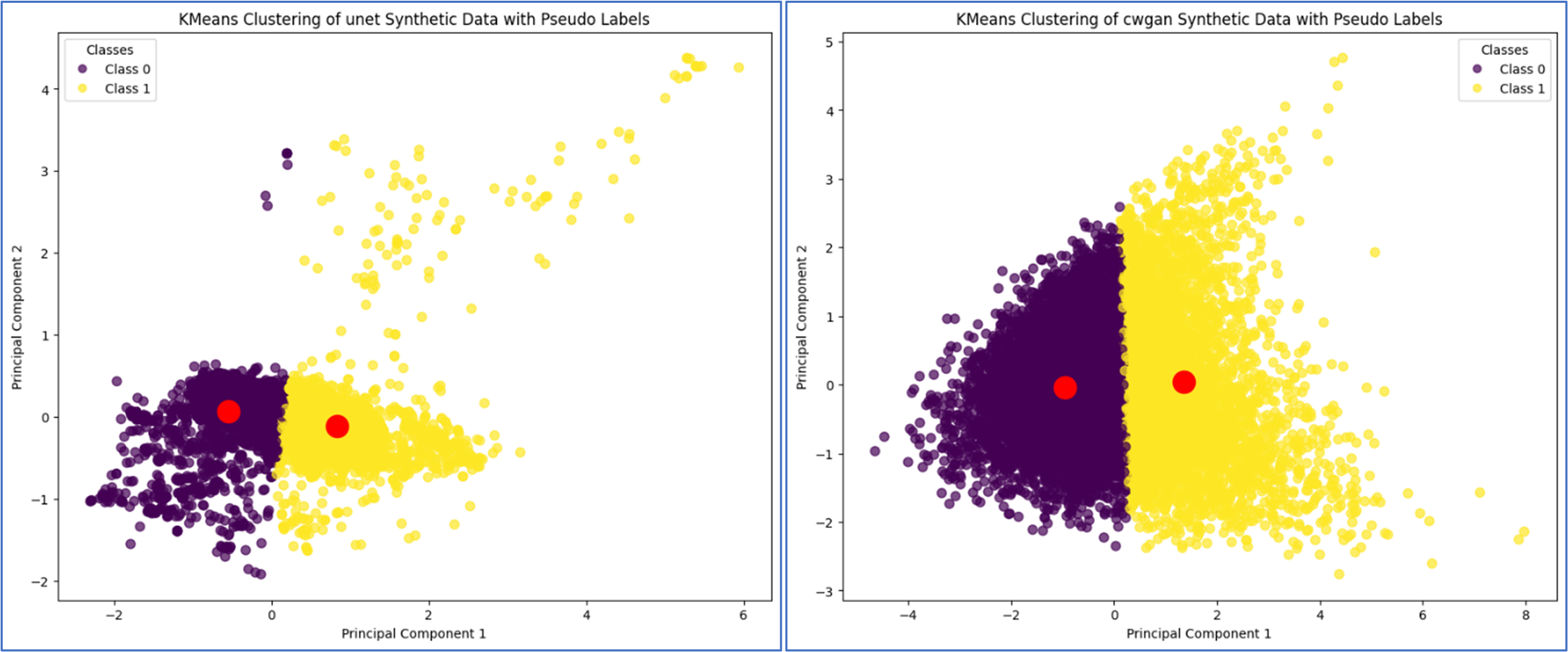}
    \caption{KMeans Clustering of U-Net and CWGAN-GP Synthetic Data with Pseudo Labels. U-Net results on the left and CWGAN-GP on the right.}
    \label{fig:kmeans_pseudo}
\end{figure}

As illustrated in Figure \ref{fig:kmeans_pseudo}, the clustering results for both U-Net and CWGAN-GP reveal a clear separation of synthetic data into Class 0 and Class 1. This process of assigning pseudo-labels enhances the practical utility of the synthetic data when integrated with real labelled datasets. By utilising these pseudo-labels, machine learning models trained on hybrid datasets, which include both real and synthetic samples, can achieve improved performance.

\subsection{XGBoost Performance on Various Data Combinations}

\begin{figure}[H]
    \centering
    \includegraphics[width=0.5\textwidth]{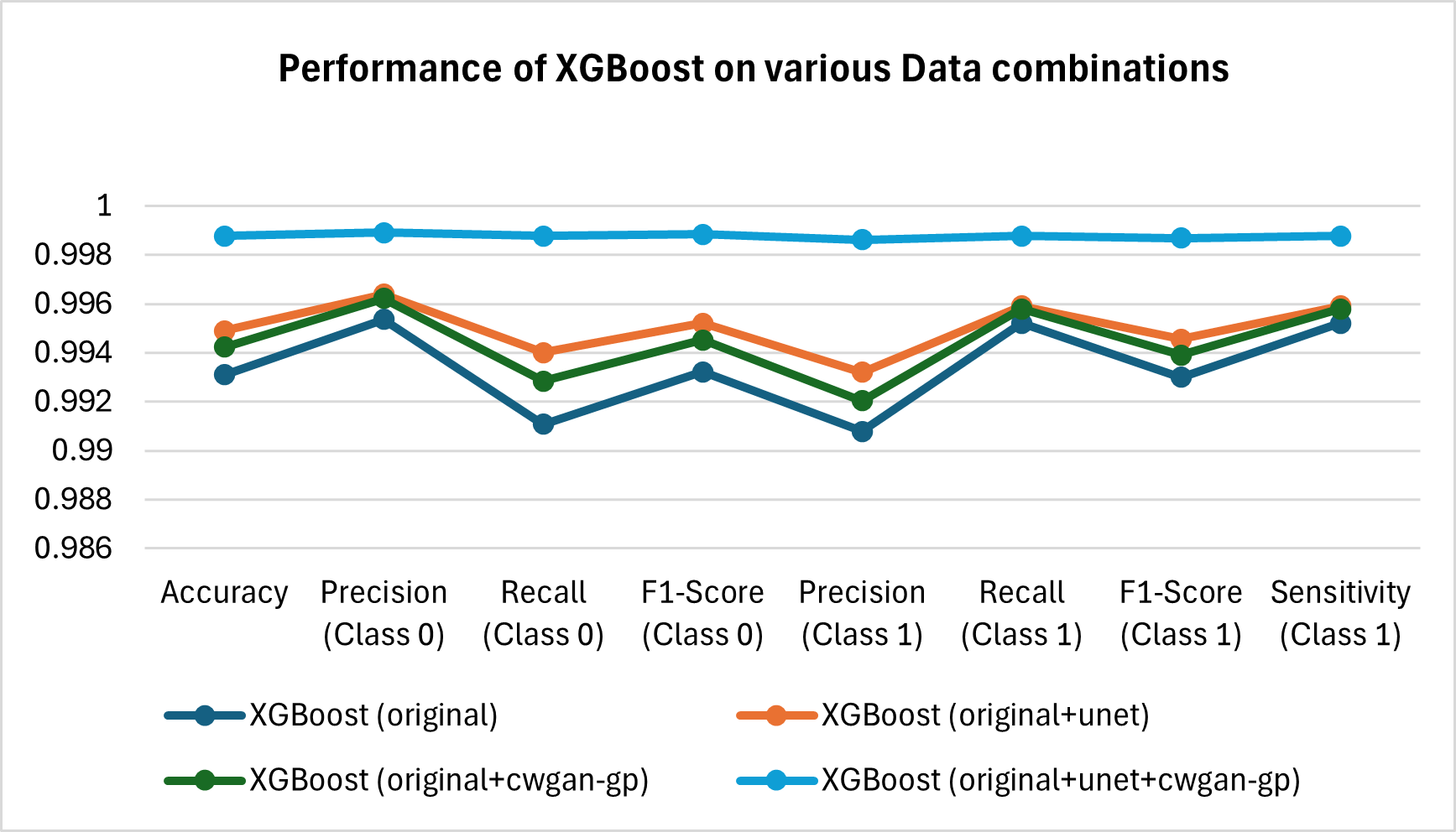}
    \caption{XGBoost Performance on Various Data Combinations}
    \label{fig:xgboost_performance}
\end{figure}

In this section, the performance of the XGBoost model trained on a combination of original data and synthetic data as generated by U-Net and CWGAN-GP models is analysed. Figure \ref{fig:xgboost_performance} illustrates the comparative results of XGBoost across various evaluation metrics.

The XGBoost model exhibited a consistently strong performance across most metrics when trained on original data. However, incorporating synthetic data from U-Net and CWGAN-GP introduced some performance variations, particularly in precision, recall, and F1-score across benign (Class 0) and malicious (Class 1) queries.

The inclusion of U-Net-generated synthetic data improved the precision and recall of the model, particularly in identifying SQL injection attacks (Class 1). Meanwhile, synthetic data as generated by CWGAN-GP provided competitive results but showed a slight reduction in sensitivity, indicating marginal difficulty in correctly identifying all malicious queries.

\subsection{Hybrid Data Proportions for XGBoost Training}

A dataset pool optimisation method was implemented by combining synthetic data from U-Net and CWGAN-GP models with real datasets in proportions to create hybrid datasets, with the goal of identifying the best balance for improving key metrics. Stratified K-Fold Cross-Validation was applied to ensure reliability in the evaluation process.

\begin{figure}[H]
    \centering
    \includegraphics[width=0.48\textwidth]{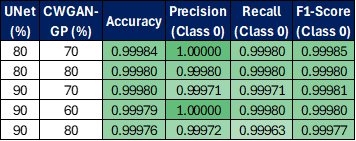}
    \caption{Performance Metrics: Class 0}
    \label{fig:class0_metrics}
\end{figure}

\begin{figure}[H]
    \centering
    \includegraphics[width=0.48\textwidth]{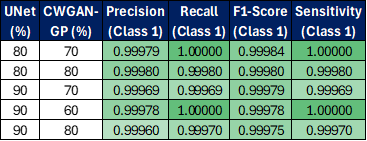}
    \caption{Performance Metrics: Class 1}
    \label{fig:class1_metrics}
\end{figure}

Among the 100 tested configurations, several exhibited a strong performance across all metrics. The top five combinations, as shown in Figures \ref{fig:class0_metrics} and \ref{fig:class1_metrics}, achieved high accuracy, precision, recall, and F1-scores. The best-performing model, utilising 80\% U-Net and 70\% CWGAN-GP synthetic data, reached an accuracy of 99.984\%, providing balanced detection for both benign (Class 0) and malicious (Class 1) SQL queries. While all top combinations showed promising results, U-Net data generally improved recall for Class 1, whereas CWGAN-GP data enhanced precision for Class 0. Ultimately, the 80\% U-Net and 70\% CWGAN-GP combination was selected for its overall balanced performance and deemed optimal for further testing and validation.

\subsection{Final XGBoost Model Performance}

With the optimal hybrid configuration established, the next step was to evaluate the final performance of the XGBoost model against the baseline model, ensuring improvements brought by synthetic data integration were robust and consistent across different datasets.

\subsubsection{Baseline Model Performance (Validation Results)}
\begin{figure}[H]
    \centering
    \fbox{\includegraphics[width=0.4\textwidth]{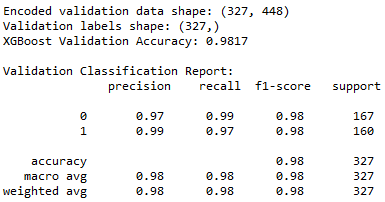}} 
    \caption{Baseline Model Classification Report}
    \label{fig:basemod}
\end{figure}

The baseline model, trained solely on the original dataset, achieved an overall accuracy of \textbf{0.9817} on the validation set. As shown in Figure \ref{fig:basemod}, the precision and recall values are well-balanced across both benign (Class 0) and malicious (Class 1) queries:

\setlength\fboxrule{0.5pt} 

The model exhibited slightly higher precision for Class 1 (malicious queries) at 0.99, while Class 0 (benign queries) showed a higher recall at 0.99. This indicates that the model was particularly effective in minimising false positives for malicious queries but showed a slight reduction in detecting all benign queries.

\setlength\fboxrule{0.5pt} 
\begin{figure}[H]
    \centering
    \fbox{\includegraphics[width=0.4\textwidth]{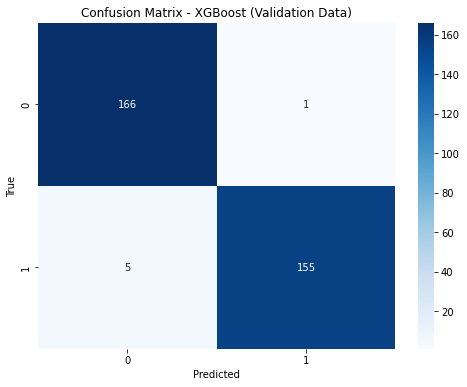}}
    \caption{Baseline Model Confusion Matrix}
    \label{fig:cmbasemod}
\end{figure}

To further illustrate the distribution of true positives, false positives, false negatives, and true negatives for both classes, the confusion matrix in Figure \ref{fig:cmbasemod} highlights the classification performance in terms of correctly and incorrectly classified instances:

While the overall performance was strong, a minor imbalance in recall for benign queries suggests potential areas for further improvement in optimising benign query detection.

\subsubsection{Final Model}

The final model was developed by leveraging the optimal data proportions of 80\% U-Net and 70\% CWGAN-GP synthetic data, as identified during the hybrid data synthesis process. This model was fine-tuned using hyperparameter optimisation, enhancing its ability to detect SQL Injection Attacks (SQLIA) with high accuracy and generalisation capability. The hyperparameters were optimised using Stratified K-Fold Cross-Validation, resulting in the following parameters:

\begin{itemize}
    \item \textbf{Best Parameters:} $\{ \text{subsample}: 1.0, \text{n\_estimators}: 500, \text{min\_child\_weight}: 1, \text{max\_depth}: 3, \text{learning\_rate}: 0.3, \text{gamma}: 0, \text{colsample\_bytree}: 0.8 \}$
\end{itemize}

\begin{figure}[H]
    \centering
    \fbox{\includegraphics[width=0.4\textwidth]{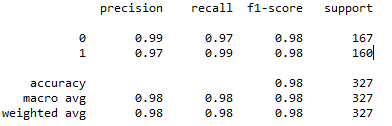}} 
    \caption{Classification Report for Final Model on Validation Data}
    \label{fig:finalmod}
\end{figure}

The final performance of the model on the validation set is depicted in Figure \ref{fig:finalmod}, where it achieved an accuracy of \textbf{0.98}. Precision, recall, and F1-scores for both Class 0 and Class 1 were well-balanced, demonstrating the efficiency of the model in detecting both benign and malicious queries.

\setlength\fboxrule{0.5pt}

\setlength\fboxrule{0.5pt}
\begin{figure}[H]
    \centering
    \fbox{\includegraphics[width=0.4\textwidth]{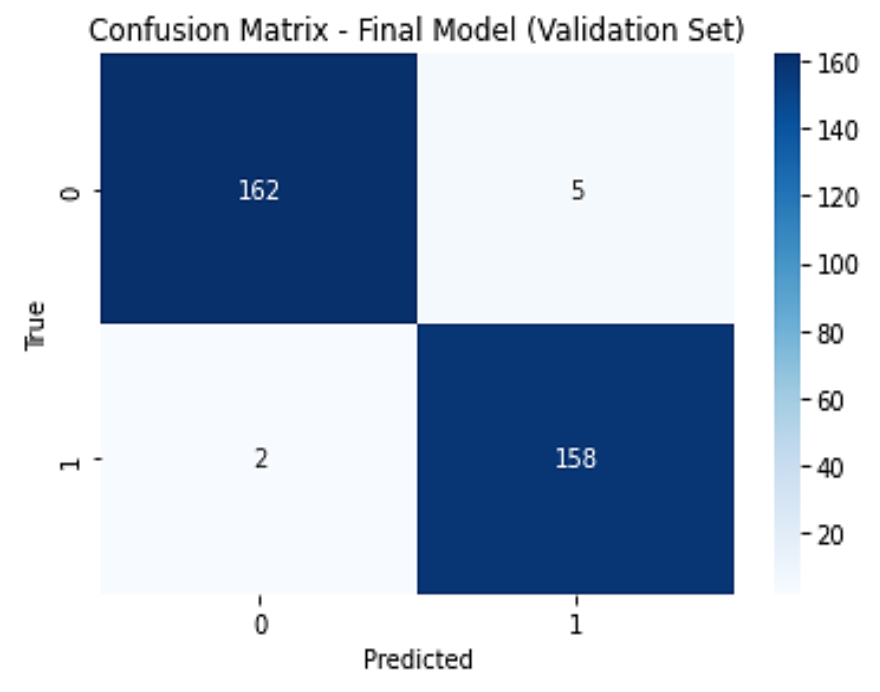}} 
    \caption{Confusion Matrix for Final Model}
    \label{fig:cm_finalmod}
\end{figure}

The confusion matrix in Figure \ref{fig:cm_finalmod} highlights the reliable classification of the model with minimal false positives and false negatives. It correctly classified 162 benign queries (Class 0) and 158 malicious queries (Class 1), showcasing the effectiveness of hyperparameter tuning in reducing classification errors.

\paragraph{\textbf{Comparison with Baseline Model}}

When compared to the baseline model, the final model exhibited a substantial improvement, particularly in the recall for Class 1 (malicious queries), which is critical for SQLIA detection. The baseline model, trained solely on real data, had a lower recall for Class 1, leading to more false negatives. By incorporating synthetic data and applying hyperparameter tuning, the final model significantly reduced these errors, providing better detection of malicious SQL queries while maintaining balanced performance for benign queries (Class 0).

\paragraph{\textbf{Cross-Validation Performance}}

\begin{figure}[H]
    \centering
    \fbox{\includegraphics[width=0.4\textwidth]{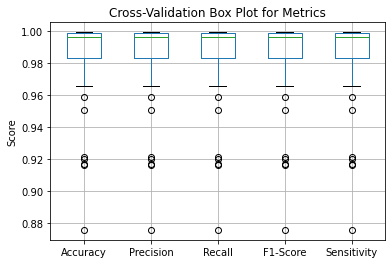}} 
    \caption{Cross-Validation Box Plot for Final Model Performance Metrics}
    \label{fig:cv_finalmod}
\end{figure}

Cross-validation results demonstrated the robustness of the final model, as depicted in the box plot (Figure \ref{fig:cv_finalmod}). Minimal variance was observed across accuracy, precision, recall, F1-score, and sensitivity metrics, further reinforcing the reliability of the model for SQL Injection detection in real-world scenarios.

\setlength\fboxrule{0.5pt}

\section{\raggedright \textbf{Conclusion}}

Overall, the final XGBoost model outperformed the baseline model, particularly in terms of recall and sensitivity for malicious queries. The integration of synthetic data, generated through U-Net and CWGAN-GP models, significantly enhanced the generalisation ability of the model. Hyperparameter tuning further optimised the performance of the model, resulting in a highly accurate and reliable solution for detecting SQLIA in real-world scenarios. This study successfully demonstrated that combining synthetic data with real data and using advanced machine learning techniques can provide an effective and robust solution for modern cybersecurity challenges.

\section{\raggedright \textbf{Limitations and Future Scope}}

While the final model demonstrated significant improvements, several limitations were identified. The CWGAN-GP model, although effective, struggled to capture the complete diversity of SQL query patterns, leading to underrepresentation of complex or rare SQL Injection Attack (SQLIA) types. Furthermore, the computational cost associated with generating synthetic data was high, making real-time deployment challenging, particularly in resource-constrained environments. Another limitation was the challenge of maintaining a balanced ratio of benign and malicious queries in the synthetic dataset, which was critical to avoid model overfitting.

Future work could focus on refining the CWGAN-GP model to better capture diverse SQL patterns and improve scalability. Exploring other GAN variants (e.g., CatGAN, SentiGAN) and hybrid models with Variational Autoencoders (VAEs) could further enhance the quality of synthetic data. Additionally, real-time detection through online learning and optimising computational efficiency using distributed systems would enable broader applications of this approach. Developing a self-learning detection system that autonomously adapts to new SQLIA types would also improve the resilience of the model to emerging threats.

\bibliographystyle{IEEEtran}
\bibliography{sqli}  

\end{document}